\documentclass[traditabstract]{aa} 
\usepackage{natbib}
\usepackage{graphicx}
\usepackage{txfonts}

\def\micron{\hbox{$\mu$m}}
\def\msun{M$_{\odot}$}

\begin{document}
   \title{ALMA reveals a warm and compact starburst around a heavily obscured supermassive black hole at $z=4.75$}
\author{
R.~Gilli\inst{1},
C.~Norman\inst{2,3},
C.~Vignali\inst{4,1},
E.~Vanzella\inst{1},
F.~Calura\inst{1},
F.~Pozzi\inst{4,1},
M.~Massardi\inst{5},
A.~Mignano\inst{5},
V.~Casasola\inst{5},
E.~Daddi\inst{6},
D.~Elbaz\inst{6},
M.~Dickinson\inst{7},
K.~Iwasawa\inst{8},
R.~Maiolino\inst{9},
M.~Brusa\inst{4,1},
F.~Vito\inst{4,1},
J.~Fritz\inst{10},
A.~Feltre\inst{11,12},
G.~Cresci\inst{1},
M.~Mignoli\inst{1},
A.~Comastri\inst{1}
\and
G.~Zamorani\inst{1}
}

\institute{
  INAF -- Osservatorio Astronomico di Bologna, via Ranzani 1, 40127 Bologna, Italy\\
    \email{roberto.gilli@oabo.inaf.it}
\and
Department of Physics and Astronomy, Johns Hopkins University, Baltimore, MD 21218, USA
\and
Space Telescope Science Institute, 3700 San Martin Drive, Baltimore, MD 21218, USA
\and
  Dipartimento di Fisica e Astronomia, Universit\`a degli Studi di Bologna, viale Berti Pichat 6/2, 40127 Bologna, Italy
\and
Istituto di Radioastronomia \& Italian ALMA Regional Centre, Via P. Gobetti 101, 40129 Bologna, Italy
\and
Laboratoire AIM, CEA/DSM-CNRS-Universit\'e Paris Diderot, Irfu/Service d'Astrophysique, CEA Saclay, Orme des Merisiers, 91191 Gif-sur-Yvette Cedex, France
\and
NOAO, 950 North Cherry Avenue, Tucson, AZ 85719, USA
\and
ICREA and Institut de Ci\`encies del Cosmos (ICC), Universitat de Barcelona (IEEC-UB), Mart\' i i Franqu\`es 1, 08028 Barcelona, Spain 
\and
Cavendish Laboratory, University of Cambridge, 19 J. J. Thomson Ave., Cambridge CB3 0HE, UK
\and
Sterrenkundig Observatorium, Universiteit Gent, Krijgslaan 281 S9, B-9000 Gent, Belgium
\and
UPMC-CNRS, UMR7095, Institut d'Astrophysique de Paris, F-75014, Paris, France
\and
ESO, Karl-Schwarzschild-Str 2, D-85748 Garching bei München, Germany
}

   \date{Received ; accepted }

 
  \abstract
{We report ALMA Cycle 0 observations at 1.3mm of LESS J033229.4-275619 (XID403), an Ultraluminous Infrared Galaxy at $z=4.75$ in the 
Chandra Deep Field South
hosting a Compton-thick QSO. The source is not resolved in our data at a resolution of $\sim$0.75 arcsec, placing an upper-limit of 2.5 kpc to 
the half-light radius of the continuum emission from heated-dust. After deconvolving for the beam size, however, we found a $\sim3\sigma$ indication of
an intrinsic source size of $0.27\pm0.08$ arcsec (Gaussian FWHM), which would correspond to $r_{half}\sim0.9\pm0.3$ kpc. We build the far-IR SED of XID403 by combining datapoints from both 
ALMA and Herschel and fit it with a modified blackbody spectrum. For the first time, we measure the dust temperature $T_d=58.5\pm5.3$ K in this system, 
which is comparable to what has been observed in other high-z submillimeter galaxies.  The measured star formation rate is 
SFR=$1020\pm150$ $M_{\odot}$ yr$^{-1}$,  in agreement with previous estimates at lower S/N. Based on the measured SFR and source size, we constrain the 
SFR surface density to be $\Sigma_{SFR}>26\;M_{\odot}$yr$^{-1}$kpc$^{-2}$ ($\sim200\;M_{\odot}$yr$^{-1}$kpc$^{-2}$ for $r_{half}\sim0.9$ kpc). 
The compactness of this starburst is comparable to what has been observed 
in other local and high-z starburst galaxies. If the gas mass measured from previous [CII] and CO(2-1) observations at low resolution 
is confined within the same dust region, assuming  $r_{half}\sim0.9\pm0.3$ kpc, this would produce
a column density of $N_H\sim0.3-1.1\times10^{24}$cm$^{-2}$ towards the central SMBH, similar to the column density of 
$\approx1.4\times10^{24}$cm$^{-2}$ measured from the X-rays. Then, in principle, if  both gas and dust were confined on sub-kpc scales, this would be 
sufficient to 
produce the observed X-ray column density without any need of a pc-scale absorber (e.g. the torus postulated by Unified Models). 
We speculate that the high compactness 
of star formation, together with the presence of a powerful AGN, likely produce an outflowing wind. This would be consistent with 
the $\sim350$ km~s$^{-1}$ velocity shift observed between the Ly$\alpha$ emission and the submm lines ([CII], CO(2-1), [NII]) and with the highly-ionized Fe 
emission line at $\sim6.9$ keV rest-frame tentatively observed in the X-ray spectrum. Finally, our observations show that, besides the mass, star formation rate
and gas depletion timescale, XID403 has also the right size to be one of the progenitors of the compact quiescent massive galaxies seen at $z\sim3$.

}

\authorrunning{R~Gilli et al.}  
\titlerunning{The warm and compact starburst around an obscured SMBH at $z=4.75$}


\maketitle

\section{Introduction}

A major effort in observational cosmology is devoted to understand the way galaxies and supermassive black holes (BHs) at their centers grow together. 
Broadly speaking, a scenario is emerging (see e.g. \citealt{hopkins08}) in which both star formation and nuclear accretion might occur in two distinct modes. 
The bulk of the galaxies and black holes would grow their mass in a secular, smooth fashion over $\sim$Gyr-long timescales \citep{daddi07,cisternas11}. 
A minor fraction of these systems, which are nonetheless the most massive and luminous, would instead assemble most of its mass during a few 
rapid bursts of activity, of the order of $\approx$0.01-0.1 Gyr  \citep{alex05}. Recent results from deep multiwavelength surveys featuring far-IR data from 
Herschel \citep{nordon10,elbaz11,rodighiero11} are confirming the distinction between the bulk of objects which are growing quietly 
(the so-called ``main sequence'') and that turbulent minority which instead is growing in bursts and is responsible for $\sim10\%$ of the cosmic star formation rate density at $z\sim2$ \citep{rodighiero11}. For this last population, mergers between gas-rich galaxies 
are believed to trigger both vigorous star formation and obscured, Eddington limited, accretion \citep{menci08}. AGN feedback is then supposed to heat 
and eventually expel the surrounding gas, quenching star formation and allowing a direct, unobscured view of the nucleus \citep{hopkins08,menci08}. 
As these systems run out of gas, quiescent galaxies are left with dormant BHs in their center. 

Ultraluminous Infrared Galaxies (ULIRGs, \citealt{sm96}) in the local Universe and bright submillimeter galaxies (SMGs) in the distant ($z\sim2-3$) Universe 
\citep{blain02} are often identified as those systems caught during their turbulent youth. 
Morphological studies show that most local ULIRGs do show merger signatures \citep{sm96}. Similarly, submillimeter galaxies often show asymmetric 
and disturbed morphologies \citep{chapman03,tacconi06,tacconi08,engel10}. 
Both populations produce stars at very high rates (SFR=100-1000 $M_{\odot}$yr$^{-1}$), 
host large reservoirs of molecular gas ($M_{gas}\sim 10^{10}\;M_{\odot}$, see \citealt{solomon05} for a review) and often hide heavily obscured, 
even Compton-thick AGN \citep{alex05,iwasawa11}. Very recently, \citet{wangbrandt13} estimated that $\approx20$\% of the SMGs in the ALMA-LABOCA survey of
the ECDFS (ALESS, \citealt{hodge13}) host X-ray detected AGN, most of which obscured with $N_H>10^{23}$cm$^{-2}$. 
The gas consumption time $(t_c = M_{gas}$/SFR) and AGN lifetime in SMGs (i.e. the time necessary 
to grow the black hole and reach the local $M_{BH}/M_{bulge}$ relation, assuming Eddington rate) are estimated to be short, $\sim 10^7$ yr. 
A particularly intriguing possibility is that during this short, possibly Compton-thick accretion phase, the BH provides maximum feedback on the surrounding 
gas through Compton heating of the galaxy ISM \citep{daddi07}. Very high column densities would indeed ensure that high-energy photons are Compton
down-scattered and then absorbed by the ISM, which can then be efficiently heated until star formation is eventually quenched. 
Heavily obscured AGN in starburst systems are therefore ideal laboratories to study  the  co-evolution of black holes with their hosts.

The fast development of submm instrumentation will soon lead to major progresses in our understanding of BH-galaxy co-evolution up to the early Universe.
The detection and identification of SMGs at $z>4$ is rapidly increasing  \citep{capak08,schinnerer08,daddi09,walter12,weiss13,vieira13}, and hyperluminous infrared systems with SFR up to $\sim3000\;M_{\odot}$yr$^{-1}$ are being discovered 
as early as $z\sim6.3$ \citep{riechers13}. The incidence and properties of obscured accreting black holes in these high-redshift systems has still to be 
properly assessed. One notable example is represented by the V-band dropout galaxy LESS~J033229.4--275619 at $z=4.75$ \citep{vanzella06}.
This object was detected in the GOODS-S field by both LABOCA and AzTEC with $f_{870\mu m}=6.3\pm1.2$ mJy and $f_{1.1mm}=3.3\pm0.5$ mJy, respectively. 
Based on these data, a total IR luminosity of 
$6\times10^{12}\;L_{\odot}$ and a SFR of about 1000 M$_{\odot}$yr$^{-1}$ were derived \citep{coppin09}. 
Observations of the CO(2-1) transition with ATCA and of the [CII]158$\mu$m line with APEX revealed large reservoirs ($10^{10}\;M_{\odot}$) of both molecular 
\citep{coppin10} and atomic gas \citep{debreuck11}. Furthermore, the 4Ms Chandra observation of the CDFS revealed the presence of a Compton-thick 
AGN in this source (also known as XID403; \citealt{xue11}) with a column density of $N_H=1.4^{+0.9}_{-0.5}\times10^{24}$cm$^{-2}$ and an intrinsic, 
de-absorbed 2-10 keV luminosity of $\sim2.5 \times 10^{44}$ erg/s \citep{gilli11}. By means of a spectral energy distribution (SED) decomposition 
technique \citep{vignali09,pozzi10}, 
the bolometric output of the AGN  was estimated  to be about half of that produced by stars \citep{gilli11}. 
XID403 therefore appears a prime interest target for further follow-up studies.

In this paper we present observations of XID403 obtained with the Atacama Large Millimeter/submillimeter Array (ALMA) in Cycle 0
and combine these data with those at other wavelengths to derive the physical properties of this system.
We summarize ALMA observations in Section 2. In Section 3 we discuss the far-IR SED obtained by combining ALMA and Herschel data. 
In Section 4 the star formation rate properties are derived. In Section 5 we combine the far-IR data with those at other wavelenghts (e.g. from the CANDELS catalog)
to derive a whole SED and perform a spectral decomposition of the stellar and nuclear components. A reanalysis of the X-ray spectrum is presented in Section 6.
Our results are discussed in Section 7 and the conclusions drawn in Section 8.
A concordance $\Lambda$CDM cosmology with $H_0=67$~km~s$^{-1}$~Mpc$^{-1}$, $\Omega_m=0.32$, $\Omega_{\Lambda}=0.68$ 
(Planck Collaboration XVI, 2013) is adopted to compute luminosities and physical sizes.

\section{ALMA observations}


We proposed continuum observation of XID403 during the first ALMA
Early Science call for proposal, aiming at significantly improving over the
existing LABOCA and Aztec detections.
The project (2011.0.00716.S; PI R. Gilli) was classified as `filler observation' and, at the end of the Cycle, got
about 23 minutes (including calibration) allocated for the Band 6
continuum observations.

The target was observed on the 21st October 2012 with a $22\times12$ m
antenna array and maximum baseline $\sim 400$m and with $4\times1.875$ GHz
spectral windows covering the frequency range 222-238 GHz, with an
averaged system temperature of $\sim 67$K. After averaging over the 7.5
GHz band the mean effective frequency for the continuum observations is
230 GHz. The time on source was $\sim 3$ min and allowed us to even overcome
the requested noise level. Neptune, PKS~0537-441, and PKS~0402-362 were observed respectively as
flux, bandpass, and phase calibrator. The data reduction was performed with CASA  (version 3.4,  \citealt{mcmullin07,petry12})
in a standard way.
Two antennas were flagged for bad system temperature and for phase drifts
between two phase calibrator observations.
The 4 spectral windows were calibrated separately and then averaged during the imaging using the MFS mode \citep{rau11}. Flux
densities were bootstrapped to the Neptune model and subsequently rescaled
according to the more recent `Butler-JPL-Horizon 2012'\footnote{Buthel et
al. 2012
https://science.nrao.edu/facilities/alma/ aboutALMA/Technology/ALMA\_Memo\_Series/alma594/abs594}
available in CASA from version 4.0 on.

\begin{figure}
\resizebox{\hsize}{!}{\includegraphics[angle=0]{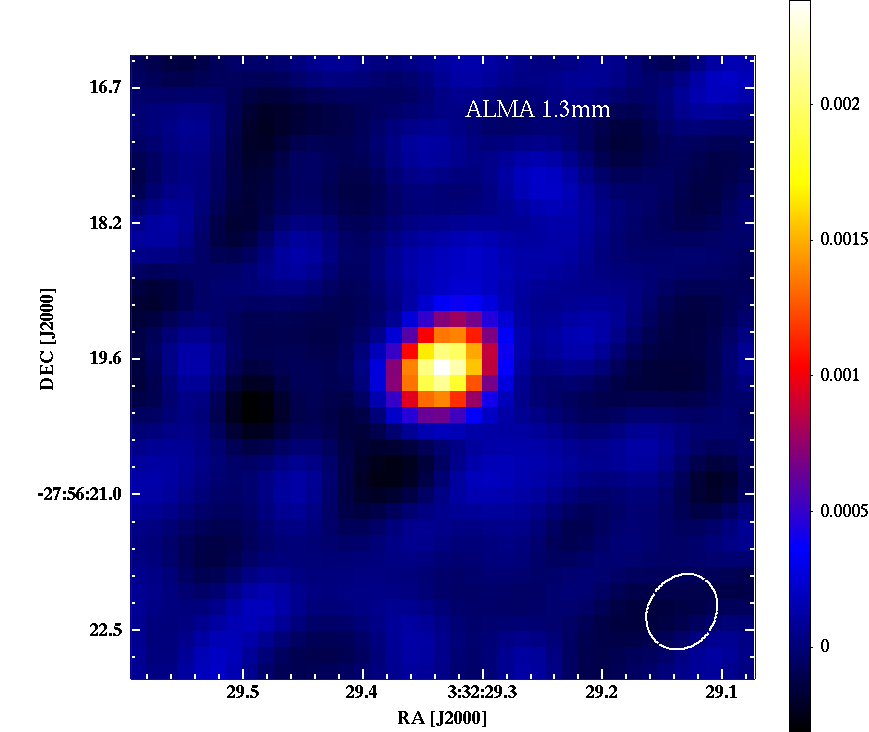}}
\caption{ALMA intensity map of the 1.3mm dust continuum emission of XID403. The $1\sigma$ rms is 0.07 mJy/beam. Units of the colorbar are
Jy/beam. The cutout is $6.5\times6.5$ arcsec, and the beam size is shown as a white ellipse ($0.84\times0.71$ arcsec FWHM) on the bottom right. 
The source is unresolved, implying a half light radius of $r_{half}<2.5$ kpc.} 
\label{almaimage}
\end{figure}

The synthesized beam FWHM of $0.84\times 0.71$ arcsec is not small
enough to image the star forming region details. However, since the
source is detected with a high signal to noise ratio, it is possible
to perform an analysis of visibilities in Fourier space and estimate
the source size on scales smaller than that of the synthesized beam.
By using the "uvmodelfit" task provided with the CASA environment, we
fitted the frequency-averaged visibilities with a single Gaussian
component with a FWHM of $0.27\pm0.08$ arcsec. This $\sim3.4\sigma$
indication of source extension has been cross checked with the package
"mapping" in the Gildas environment, which provided similar results.
We note, however, that this measurement needs to be confirmed by data at even higher resolution, 
since at our longest baselines the sampling of the uv plane is sparse and the 
flux measurement could be decreased by phase noise.
The rms achieved in our image at 1.305mm is 0.07 mJy, and we measured a flux density at the peak of $S_{1305}=2.47$ mJy.
The ALMA Band 6 image is shown in Fig.~\ref{almaimage}.

\begin{table}
\caption{Summary of far-IR photometric datapoints}
\begin{tabular}{rrcccl}
\hline \hline
$\lambda_{obs}$& $\lambda_{rest}$& $S_{\nu}$& Beam& Observatory& Refs\\
($\mu$m)& ($\mu$m)& (mJy)& (arcsec$^2$)& & \\ 
\hline
70&        12&            $<0.9^a$&      $6.5\times6.5$&      Herschel/PACS& 1\\
100&      17&            $<0.6^a$&      $7.4\times7.4$&      Herschel/PACS& 1\\
160&      28&           $<1.3^a$& $11.3\times11.3$&      Herschel/PACS& 1\\
250&      43&      $4.1\pm1.9$&        $18\times17$&     Herschel/SPIRE& 2\\
350&      61&               $<12^a$&        $25\times23$&     Herschel/SPIRE& 3\\
500&      87&               $<15^a$&        $37\times33$&     Herschel/SPIRE& 3\\
872&    152&      $6.1\pm0.5 $&      $2.1\times1.1$&         ALMA band7& 4\\
1181&  205&        $3.5\pm0.1$&      $1.7\times1.5$&         ALMA band6& 5\\
1305&  227&  $2.47\pm0.07$&  $0.84\times0.71$&        ALMA band6& 6\\
7500&  1304&            $<0.04^a$&           $12\times9$&                ATCA& 7\\
\hline                                             
\end{tabular}
\label{tab1}
Notes: $^a3\sigma$ upper limit. References: 1) \citet{magnelli13}; 2) Daddi et al. in preparation; 3) \citet{magnelli12}; 4) \citet{hodge13}; 5) \citet{nagao12}; 
6) this work; 7) \citet{coppin10}.
\end{table}

\section{Far-IR SED: dust mass and temperature}

Besides our data, there are two more continuum observations performed with ALMA on this source. 
The first was obtained by \citet{nagao12} at poorer resolution as a byproduct of their investigation of source metallicity
through the [N II]205$\mu$m line: they measured a continuum flux density at $\sim$1.2 mm of $S_{1181}=3.5\pm0.1$ mJy. The other measurement
was performed by \citet{hodge13} within a band 7 survey of the SMGs detected by LABOCA in the E-CDFS (XID403 is their ALESS 73.1): they measured
 $S_{872}=6.1\pm0.5$ mJy. These values are in excellent agreement with the flux densities derived from the $5\sigma$ detections
by AzTEC and LABOCA at 1.1mm and 870$\mu$m, respectively. At shorter wavelengths, faint emission can be seen at the position of XID403 
in the Herschel/SPIRE map at 250$\mu$m, for which a flux measurement was derived (at $\sim2\sigma$; Daddi et al. in preparation). 
Only upper limits to the source flux are available from the Herschel/PACS data (70, 100, 160$\mu$m) or Herschel/SPIRE data at 350 and 500$\mu$m \citep{magnelli12}.
For PACS observations, we used the $3\sigma$ limits available for the deepest part of the PEP-GOODS-Herschel combined map \citet{magnelli13}. Looser limits
would be more appropriate at the position of XID403, but they would not have any impact on the derived best-fit SED (see below). 
For observations at 350$\mu$m and 500$\mu$m, we simply considered the $3\sigma$ flux upper limits as derived in the SPIRE 24$\mu$m-prior catalogs 
($3\sigma$ confusion limits are slightly above these values \citep{magnelli12}. Recently, \citet{swinbank13} produced fully deblended SPIRE photometry of ALMA sources in the ECDFS, 
deriving $3\sigma$ limits of $\sim 7$, $\sim$8, and $\sim10$ mJy for XID403 (see their Fig. 4), which are consistent with the numbers 
we quote in Table 1 and with our best-fit SED (see below).
An upper limit to the continuum flux at longer wavelengths ($\sim 7.5$mm) is also available from ATCA observations \citep{coppin10}. 
A summary of the available far-IR data is shown in Table~1.  

We fit these data with a general modified blackbody model of the form \citep{rangwala11}:

\begin{equation}
S_{\nu_{obs}} \propto B_{\nu}(T) (1 - \mathrm{e}^{-\tau_{\nu}}), 
\end{equation} 
where\\
\begin{equation}
\tau_{\nu} ={\left(\frac{\nu}{\nu_{0}}\right)^{\beta}}  \,\,\, {\rm and} \quad   B_{\nu}\,(T) = \frac{2h\nu^{3}}{c^2}\frac{1}{\mathrm{e}^{\frac{h\nu}{kT}} - 1}.
\end{equation}

\begin{figure}
\resizebox{\hsize}{!}{\includegraphics[angle=270]{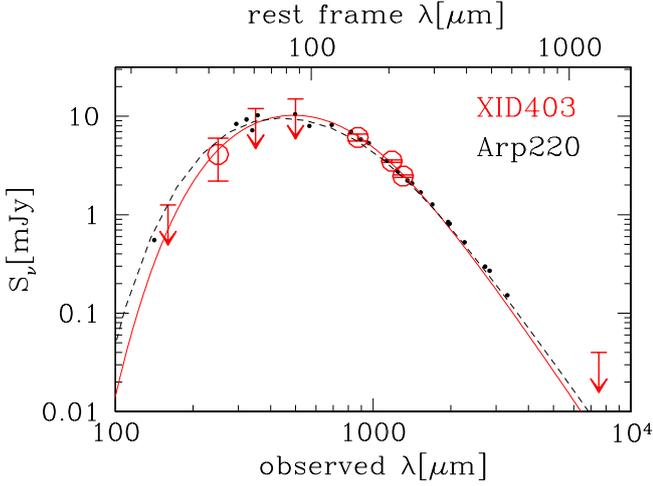}}
\caption{Far-IR spectral energy distribution (SED) of XID403 (big red symbols and solid line) compared to that of the local starburst galaxy Arp220 
(small black dots and dashed line; \citealt{rangwala11}) shifted to z=4.75 and renormalized. The far-IR photometry of XID403 is reported in Table~1. 
The far-IR SED was fitted with a general 
modified black-body by fixing the emissivity index to $\beta=2$ and the rest-wavelength below which the dust is optically thick to $\lambda_0=200\mu$m,
as commonly observed in high-z SMGs \citep{conley11,fu12,riechers13} and also in Arp220. 
The best-fit temperature is $T_{\rm d}=58.5\pm5.3$K. The best-fit dust mass  is $M_{\rm d}=4.9\pm0.7\times 10^8\;M_{\odot}$ (see text for details).}
\label{firsed}
\end{figure}

In the above equations $\nu_{obs}$ and $\nu$ are the observed and rest frequency, respectively [$\nu=\nu_{obs}(1+z)$], $\tau_\nu$ is the dust
optical depth, and $B_{\nu}\,(T)$ is the Planck function.
Because of the limited number of photometric datapoints, we fixed the dust emissivity index to $\beta=2$ and the rest-wavelength at
which the dust becomes optically thick (i.e. $\tau_{\nu} =1$) to $\lambda_0=c/\nu_0=200\mu$m (i.e. $\nu_0$=1.5 THz). 
These are consistent with the best fit values observed 
in distant SMGs with densely sampled SEDs \citep{conley11,fu12,riechers13} and with what has been observed 
in the local ULIRG Arp220 \citep{rangwala11}. We note that most of the FIR datapoints available for XID403 are at $lambda_{rest}\lesssim200\;\mu$m,
where the dust is likely optically thick. Therefore, the optically-thin grey-body approximation 
$S_{\nu_{obs}} \propto B_{\nu}\tau_{\nu}$, which is often used in the literature to fit FIR SEDs, cannot be utilized in this case, since it would severely 
underestimate the dust temperature (see e.g. \citealt{conley11,fu12}). A simple $\chi^2$ fit to the FIR SED of XID403 returns an acceptable
statistics with a best fit temperature of $T_{\rm d}=58.5\pm5.3$K (see Fig.~\ref{firsed}), which is similar to what is measured with the same grey-body 
model in Arp220 ($66.7\pm0.3$K) and in high-z SMGs with accurate SED determination.
Once the best fit temperature and model normalization have been found, we determined the dust mass which is responsible for the total FIR emission
by considering the model flux density at 1.3mm, where the dust is optically thin ($\lambda_{rest}=227\mu{\rm m}>\lambda_0$) and the residual wrt 
to the ALMA data is minimal. We then used the standard relation valid for the optically thin case (e.g.  \citealt{coppin10,greve12}):
\begin{equation}
M_{\rm d} = \frac{D_{\rm L}^2 S_{\nu_{obs}} }{(1+z) \kappa_{\nu}B_{\nu}(T_{\rm d})},
\end{equation}

where $S_{\nu_{obs}}$ is the observed flux density at 1.3mm and $\kappa_{\nu}$ and $B_{\nu}(T_{\rm d})$ are computed at $\nu=\nu_{obs}(1+z)$.
We adopt $\kappa_{\nu} = 4.0 \left (\nu/1.2{\rm THz} \right)^{\beta}{\rm cm^2\,g^{-1}}$
with $\beta=2.0$. These prescriptions for the absorption cross section $\kappa_{\nu}$ and the emissivity index $\beta$ are consistent with those derived 
from full dust models which consider a distribution of dust grains and a range of interstellar radiation field densities (e.g. \citealt{draine07}). 
The dust masses determined with Eq.3 using these prescriptions are in excellent agreement with those determined by full dust models \citep{bianchi13}. 
For XID403 we measured $M_{\rm d} =4.9\pm0.7\times10^8\;M_{\odot}$. The large dust content of this object is comparable to the ones derived in starbursting 
QSO hosts at similar or even higher redshifts \citep{wang13,calura13}.

\begin{figure}
\resizebox{\hsize}{!}{\includegraphics[angle=0]{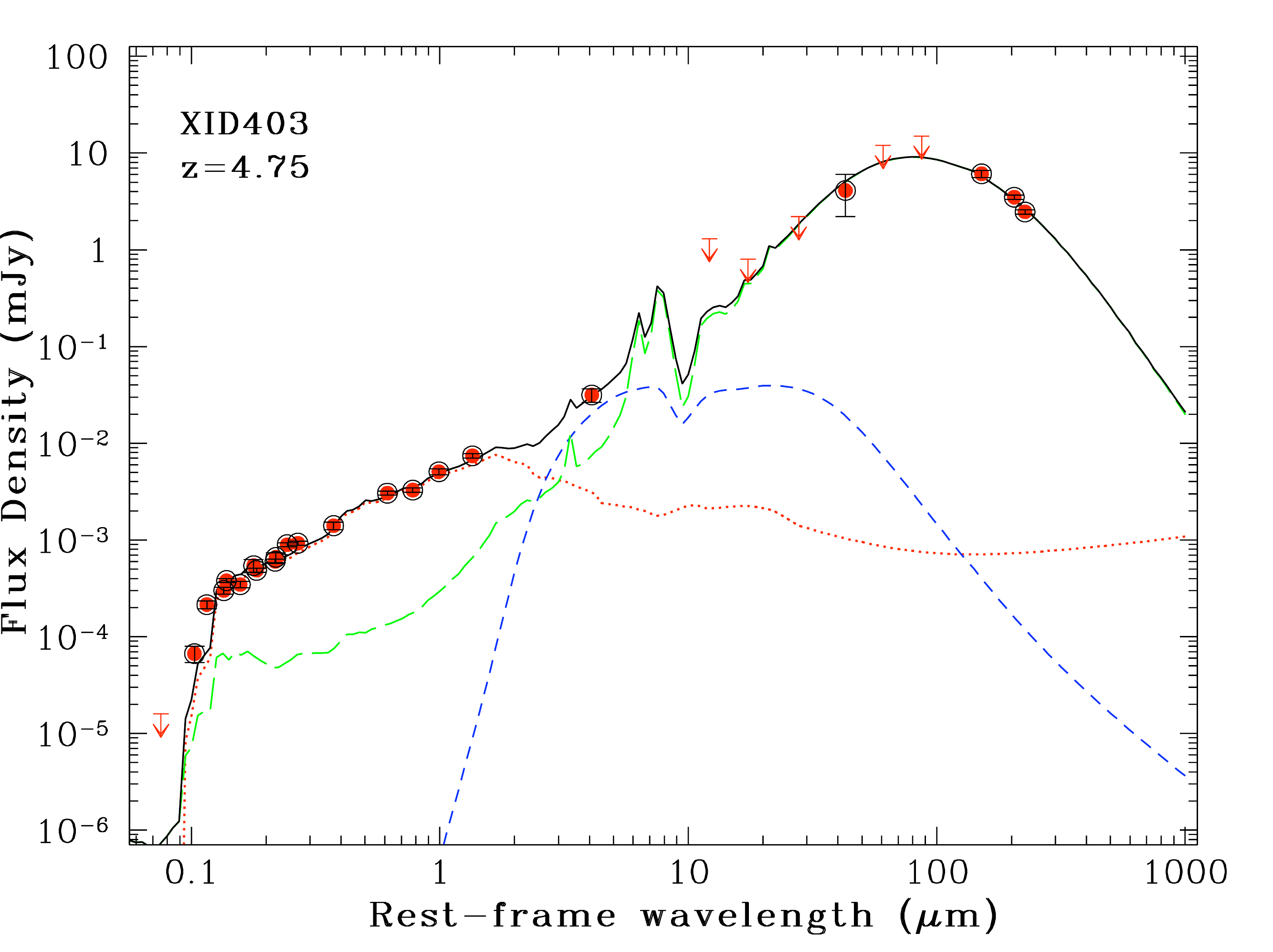}}
\caption{Rest-frame SED of XID403 and its spectral decomposition. 
Photometric datapoints and 3$\sigma$ upper limits are marked by red filled 
points and downward-pointing arrows, respectively. 
The total SED (black solid line) is the summed contribution of the stellar 
(red dotted line), AGN (blue short-dashed line), and star-forming (green 
long-dashed line) components. See text for details.}
\label{sed_ds}
\end{figure}

\section{Star formation rate and compactness}

By integrating the best-fit modified greybody spectrum we obtained a far-IR (42.5-122.5$\mu$m) luminosity of 
$L_{FIR}=4.2\pm0.6\times10^{12}\;L_{\odot}$ and a total-IR (8-1000$\mu$m) luminosity of 
$L_{IR}=5.9\pm0.9\times10^{12}\;L_{\odot}$, where the 15\% quoted uncertainties just take into account the statistical uncertainties
on the greybody model normalization. These values are in excellent agreement with those previously derived by \citet{coppin09}
and \citet{gilli11} and have significantly smaller uncertainties.
By using the \citet{kennicutt98} relation SFR$(M_{\odot}/yr) = 1.74\times 10^{-10} L_{IR}(L_{\odot})$  we derived a
star formation rate of SFR$=1020\pm150\;M_{\odot}$yr$^{-1}$. We note that this value has been obtained by the integration
of the single greybody component, and therefore is not contaminated by AGN emission, which instead contributes
significantly to the $total$ SED at $\lambda_{rest}\lesssim20\mu$m (see \citealt{gilli11} and Fig.~3). 
Based on the ALMA image at 1.3mm, where the source emission is spatially unresolved on sub-arcsec scales, 
(beam FWHM$\sim$0.75 arcsec) we can derive an upper limit of $r_{half}<2.5$kpc to the half-light radius of the dust continuum emission 
($r_{half}=0.9\pm0.3$ kpc based on the analysis of visibilities)\footnote{
One arcsec is 6.6kpc at $z=4.75$; we also recall that, for a Gaussian beam, $r_{half}=$FWHM/2.}. The observed star formation hence is produced
within a kpc-scale region, and a lower limit to the rate of star formation per unit area  $\Sigma_{SFR}=(SFR/2)/(\pi r_{half}^2)>26\;M_{\odot}$yr$^{-1}$kpc$^{-2}$ 
can be obtained ($\Sigma_{SFR}=200\;M_{\odot}$yr$^{-1}$kpc$^{-2}$ for $r_{half}=0.9$ kpc). Highly-starforming galaxies (LIRGs/ULIRGs) both in the local and distant Universe are usually making their stars in compact bursts 
with $\Sigma_{SFR}\sim10-100\;M_{\odot}$yr$^{-1}$kpc$^{-2}$ (see e.g. \citealt{genzel10} and references therein). \citet{hodge13}
derived a median SFR surface density of $>14\;M_{\odot}$yr$^{-1}$kpc$^{-2}$ for their sample of SMGs observed by ALMA in the ECDFS at median $z\sim2.2$. 
The value we measured for XID403 is in line with these findings. 

\section{SED decomposition}

The CDFS features the deepest multiwavelength database available in a given patch of the sky. Very recently the CANDELS GOODS-South
catalog has been released \citep{guo13} which provides optical to near-IR photometry information for about 35,000 H-band detected sources
(HST/WFC3 F160W band) in the GOODS portion of the CDFS. XID403 is detected at $1.6\mu$m with a flux of $0.926\pm0.023\;\mu$Jy (23.89 AB mag).
We retrieved all the photometry available for XID403 from the CANDELS catalog, including HST optical/near-IR imaging with 
ACS and WFC3 (from 0.4 to 1.6 $\mu$m), VLT K-band imaging (HAWK-I and ISAAC) and Spitzer/IRAC near-IR imaging (from 3.6 to 8.0 $\mu$m). 
We added to those data Spitzer/MIPS photometry at 24$\mu$m and the FIR data from Herschel and ALMA described in the previous Sections. 
A complete and extremely well sampled SED is then obtained, which is shown in Fig.~\ref{sed_ds}.

The optical to far-infrared (millimeter) data of XID403 were modelled using 
the spectral energy distribution (SED) fitting code originally developed by 
\citet{fritz06} and recently updated by \citet{feltre13}. This code 
allows the user to fit broad-band data using three components: 
i) a stellar component for the host galaxy, whose contribution is dominant in the 
optical/near-IR bands; ii) an AGN component due to nuclear emission 
(mostly UV) reprocessed by hot dust, peaking at a few tens of microns; iii)
a star-forming component, which accounts for the far-IR blackbody-like thermal 
bump (see \citealt{pozzi10,pozzi12} and \citealt{vignali09,vignali11} for recent applications of this code).

The stellar component (red dotted line in Fig.~\ref{sed_ds}) comprises a set 
of simple stellar population (SSP) spectra of solar metallicity and ages up 
to 0.6~Gyr, i.e., the time elapsed between z$_{\rm form}$=8 (the redshift here
assumed for the stars to form) and $z=4.75$ (the source redshift). 
Extinction to the stellar component is modelled using a \citet{calzetti00}
attenuation law. 

We  modeled the AGN  component (blue short-dashed line in Fig.~\ref{sed_ds}) using 
a smooth torus model with a ``flared disc''  geometry (see \citealt{fritz06}). Although high-resolution 
mid-IR observations of nearby AGN have suggested that a clumpy configuration for the obscuring torus 
is probably more likely (e.g., \citealt{jaffe04,tristram07,burtscher13}), smooth and clumpy 
torus models provide equally good descriptions of the AGN infrared SEDs (as extensively discussed by \citealt{feltre12}).

Finally, the far-IR emission has been reproduced using dusty SSPs (comparably good results are obtained using empirical starburst FIR templates). 
The SSP fit is shown as a 
long-dashed line in Fig.~\ref{sed_ds}). These dusty SSPs share the same star formation history of those 
used to fit the optical-UV part of the spectrum: their emission 
appears strongly extinguished in the optical  (see Fig.~\ref{sed_ds}),
but they are able to reproduce the whole far-IR thermal bump. 
%

Overall, the SED fitting to the available datapoints is extremely good 
(black solid line in Fig.~\ref{sed_ds}) and allows us to estimate many of the relevant physical 
parameters of both the host galaxy (e.g. stellar mass, optical extinction, SFR) and the AGN (e.g. bolometric luminosity). 
The large amount of information at optical/near-IR wavelengths provides a reliable estimate of the stellar mass $M_*\sim1.1\times10^{11}$ \msun 
( the average value measured for ALESS SMGs is $8\pm1\times10^{10}\;M_{\odot}$; \citealt{simpson13}), where
the estimated uncertainties on $M_*$ are within 50\% and are dominated by systematics, e.g. by the choice of the IMF (we used a Salpeter IMF). 
Significant extinction to the host galaxy emission is also required 
[E(B$-$V)=0.33]. 
For what concerns the AGN, its contribution mostly relies on the 
{\it Spitzer}/MIPS photometric datapoint at 24\micron\ (see also 
\citealt{coppin09}). This does not allow us to constrain 
the geometric AGN parameters (e.g., the opening angle of the torus). However, 
the source bolometric accretion luminosity appears to be well determined to $\sim10^{46}$ erg~s$^{-1}$ ($2.6\times10^{12}\;L_{\odot}$).
This value is in excellent agreement with what is derived from the 2-10 keV intrinsic luminosity
assuming appropriate bolometric corrections \citep{lusso12} and is about half of the total luminosity produced
by stellar processes ($L_{IR}=5.9\times 10^{12}\;L_{\odot}$).

%
%
As evident in Fig.~\ref{sed_ds}, the AGN contributes 2--3\% of the 
8--1000\micron\ emission. Once corrected for this contribution, the far-IR 
luminosity can be used to derive a star-formation rate of $\sim$1040~\msun/yr,
in excellent agreement with what has been estimated in Section 4. 
This value is much larger than the one obtained by fitting the optical/near-IR SED ($\sim$290~\msun/yr). As discussed in Section 7.2, 
such a discrepancy is often found when fitting the broad band SEDs of LIRGs/ULIRGs and can be explained, at least to zeroth order, in terms of most star-formation 
occurring in heavily obscured, dust-embedded environments (as also assumed in dusty SSP models; \citealt{dacunha08}). 

\begin{figure}
\resizebox{\hsize}{!}{\includegraphics[angle=270]{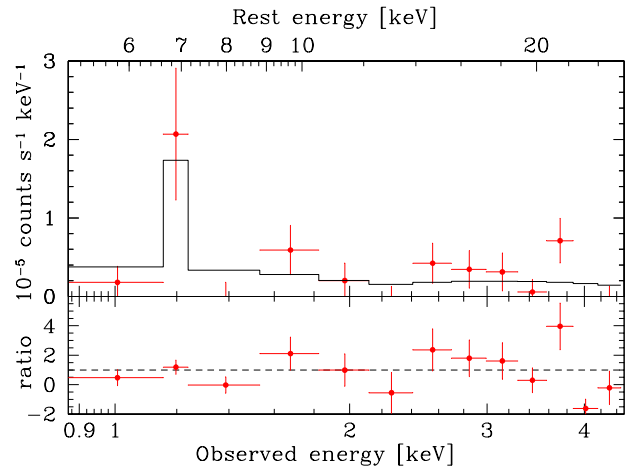}}
\caption{4Ms Chandra X-ray spectrum of XID403 and ratio to the best fit model (see text for details). A Fe emission line at $\sim6.9$ keV rest-frame is detected
at $\sim2\sigma$.}
\label{xray}
\end{figure}

\section{X-ray emission: photons from highly-ionized iron?}

As reported by \citet{gilli11}, the hard X-ray spectrum of XID403 strongly points to heavy, Compton-thick absorption.
The best-fit absorption column density was found to be $1.4^{+0.9}_{-0.5}\times10^{24}$cm$^{-2}$. Similar values were derived by 
Vito et al. (2013; $1.8^{+1.5}_{-0.8}\times10^{24}$cm$^{-2}$) and Wang et al. (2013b; $\approx 0.8\times10^{24}$cm$^{-2}$). However, even higher columns could not
be ruled out. Indeed, when $N_H\gtrsim10^{25}$cm$^{-2}$, the direct X-ray nuclear radiation is completely blocked, and only
photons scattered by matter off the line of sight can be detected. In the case of XID403, a simple Compton reflection continuum from 
cold gas surrounding the nucleus provided an equally good fit to the X-ray spectrum. 
In \citet{gilli11} we did not detect any iron emission line at 6.4 keV rest-frame. This prominent line (equivalent width EW$\sim1-2$keV) is commonly 
observed in Compton-thick AGN (e.g. \citealt{comastri04}) and usually interpreted as a fluorescence line associated to low-ionized iron atoms (up to Fe~{\sc xvii}) 
in the cold reflector. The upper limit we could place on the line EW, however, was very loose (EW$<4.3$keV at 90\% confidence level) 
and did not contrast with the interpretation of a heavily obscured nucleus. 
Here we have re-analyzed the 4Ms Chandra X-ray spectrum of XID403 and performed a more detailed search of iron emission features. 
In particular, we were prompted by the results of \citet{iwasawa12cosmos}, who found evidence of high-ionization lines from 
Fe~{\sc xxv} at 6.70 keV and Fe~{\sc xxvi}
at 6.97 keV in the stacked spectrum of  33 obscured AGN hosted by ULIRGs in the COSMOS field. 
The details of the data reduction and X-ray spectral analysis of XID403 are presented in \citet{gilli11}. 
We now fit that spectrum with a pure reflection continuum (pexrav model in XSPEC) plus an iron line at a centroid
energy which is left free to vary. We used the Cash statistic to estimate the best fit parameters. Errors are quoted at $1\sigma$ confidence level ($\Delta C=1$). 
We found a $\sim2\sigma$ evidence for a line at $\sim1.19$keV, i.e. $\sim$6.9 keV rest-energy ($6.88\pm0.16$ keV; see Fig.\ref{xray}), which can be intepreted 
as emission from highy ionized iron. The same feature is confirmed when we consider the Chandra spectrum obtained by \citet{vito13} who performed a 
completely independent spectral extraction and background subtraction. As a safety check, we verified that no significant emission features at that energy 
are present in the Chandra background spectrum. 
The rest-frame line EW is $2.8_{-1.4}^{+1.7}$keV: indeed, given the low S/N of the spectrum and the redshift dimming of the observed equivalent width
[EW$_{obs}$=EW$_{rest}/(1+z)$], only extreme features can be detected with some confidence. 
The measured EW is very large for what is commonly seen in X-ray spectra but not unreasonable (see \citealt{matt96} for a theoretical perspective). 
In particular, there is mounting evidence that ULIRGs and SMGs both in the local and distant Universe feature
He-like and H-like Fe lines at 6.7 and 6.9 keV with large, $\sim1-2$ keV, equivalent widths (e.g. \citealt{iwasawa09,lindner12}). 
In particular, such lines with EW up to $\sim2$ keV have been detected in powerful starforming objects hosting heavily obscured AGN
\citep{iwasawa05,nandra07,nardini11,jia12}. The tentative detection in XID403 of an emission feature from highly-ionized iron  
would be in line with these findings. We will return on the Fe line detection and its interpretation in the Discussion.

\begin{table}
\caption{Summary of the physical properties of XID403}
\begin{tabular}{cccc}
\hline \hline
Quantity& Value&  Units& Reference\\

\hline
$L_{FIR}$&$4.2\pm 0.6$& $10^{12} L_{\odot}$& this work\\
$L_{IR}$&$5.9\pm 0.9$& $10^{12} L_{\odot}$& this work\\
$L^{bol}_{AGN}$& $2.6$& $10^{12} L_{\odot}$& this work\\
SFR& $1020\pm150$& $M_{\odot}$yr$^{-1}$& this work\\
$M_{dust}$& $4.9\pm0.7$& $10^{8} M_{\odot}$&  this work\\
$T_{dust}$& $58.5\pm5.3$& K& this work\\
$r_{half}$& $<2.5$& kpc& this work\\
$\Sigma_{SFR}$&$>26$& $M_{\odot}$yr$^{-1}$kpc$^{-2}$& this work\\
$M_{HI}$& $1.0\pm0.3$& $10^{10} M_{\odot}$& De Breuck et al. (2011)\\
$M_{H_2}$& $1.6\pm0.3$& $10^{10} M_{\odot}$& Coppin et al. (2010)\\
$\Sigma_{H1+H_2}$&$>6.6$ & $10^{8} M_{\odot}$kpc$^{-2}$& this work\\
$N_{HI+H_2}$&$>0.1$& $10^{24}$cm$^{-2}$& this work\\
$N_H^{X-ray}$&$1.4^{+0.9}_{-0.5}$ & $10^{24}$cm$^{-2}$& Gilli et al. (2011)\\
$M_{dyn}sin^2_i$& $1.2\pm0.6$& $10^{10} M_{\odot}$& Coppin et al. (2010)\\
$M_*$& $11$& $10^{10}M_{\odot}$& this work\\
$Z$& $1.0^{+1.0}_{-0.6}$& $Z_{\odot}$& Nagao et al. (2012)\\
\hline                                             
\end{tabular}

Notes: The dynamical mass estimated by \citet{coppin10} assumes a disk-like geometry where
$i$ is the inclination angle of the disk axis to the line of sight.

\label{tab1}
\end{table}

\section{Discussion}

\subsection{Size of the system}

The sub-arcsec ALMA observations reported in this work constrain the FIR dust continuum to be emitted within a half-light radius of $r_{half}<2.5$ kpc 
from the central SMBH (and also suggest at the $\sim3\sigma$ level that $r_{half}$ could be around 0.9 kpc).
Observations of high-z QSOs \citep{wang13} show that both CO and [CII] emission are co-spatial with FIR emission. Furthermore, \citet{riechers11} showed that, 
as opposed to high-z SMGs, high-z QSOs do not show evidence of extended, low-excitation molecular gas components.
Then, under the reasonable assumption that both molecular and atomic gas are co-spatial with dust also in XID403, we obtain that, if half of the total gas mass 
$M_{gas} = M_{H_2} + M_{HI} = 2.6\times10^{10}\,M_{\odot}$ \citep{coppin10,debreuck11} is confined within 2.5 kpc from the nucleus, this 
would produce a column density of $N_H>0.1\times10^{24}$cm$^{-2}$ towards it.\footnote{Based on a recent estimate from ALMA data
(De Breuck et al. priv. comm.), the [CII] luminosity and hence $M_{HI}$ derived in \citet{debreuck11} are likely overestimated by a factor of $\sim2$. This would decrease 
all our estimates of total gas densities by $\sim20\%$, without affecting significantly our results} . For $r_{half}=0.9\pm0.3$ kpc, we estimate 
$N_H=0.3-1.1\times10^{24}$ cm$^{-2}$,
which is consistent with what is inferred from the X-ray spectrum and shows that, in principle, 
absorption on galaxy scales could be sufficient to hide the nucleus without invoking any additional pc-scale obscuring matter (the estimated column density
would instead be overestimated if the gas reservoir is more extended than dust; see e.g. \citealt{swinbank12} and references therein). Recent numerical 
simulations indeed show that high redshift galaxies hosting dense and thick disks of gas on kpc scales may provide heavy absorption (up to Compton-thick) 
towards their central black hole \citep{bournaud11}, and a physical link between nuclear absorption and host gas content has been suggested even for AGN at $z<1$ \citep{juneau13}.
We note that the column density derived from the X-ray spectrum was obtained assuming a solar metallicity for the obscuring medium. This is consistent 
with what has been measured by \citet{nagao12} for the metallicity of the gas in the circumnuclear starburst.
Nevertheless, we cannot rule out the presence of a compact inner absorber around the nucleus. The presence of a dense medium
on pc scales from the nucleus is indeed revealed by the presence of hot, AGN-heated dust at 4$\mu$m rest-frame (see Fig.~\ref{sed_ds}). However, 
since both the covering factor and the inclination along the line of sight of this compact medium cannot be constrained,  it is not clear how much of the 
nuclear obscuration it produces.


The HST observations in the UV/optical rest-frame can be used to get another estimate of the size of the system, although it is likely that UV light 
is not tracing the entire dimensions of the system as most star formation is occurring in dust-obscured regions (see Section 5). The emission in the
HST/ACS z-band  is unresolved at 0.1 arcsec resolution (FWHM), pointing to a half-light radius of stellar emission of $<$0.3 kpc at $\sim 1600\AA$ rest-frame 
\citep{coppin09,gilli11}.  In the HST/WFC3 H-band, the source is detected with a stellarity parameter of 0.9 \citep{guo13}. That is, the source 
could be marginally resolved at a resolution of 0.3 arcsec at $\sim2800\AA$ rest-frame, which would give a half-light radius  of $\sim1$ kpc for 
stellar emission, similar to what is found for dust emission.
Any contribution of the AGN to the optical/UV emission, e.g. through scattered light, cannot be quantified precisely. However, as opposed to 
\citet{gilli11}, we suggest here that most of the optical/UV light is produced by stars. First, the broad band SED obtained from the CANDELS
catalog seems very smooth and a simple galaxy template provides an excellent fit  without requiring any additional 
component (see Fig.~\ref{sed_ds}). Second, extremely compact morphologies with half-light radii as small as $\sim0.1$ kpc have been recently observed 
in highly star forming systems without AGN that reach star formation rate densities of $\sim3000\;M_{\odot}$yr$^{-1}$kpc$^{-2}$ \citep{diamond12}. 
Third, the rest-frame UV spectrum of XID403 closely resembles those of local compact starbursts (see Section 7.4).
We have therefore some indications that, if both stars, gas and dust were co-spatial,  the system size could realistically be $\sim0.9$ kpc 
and as small as 0.3 kpc.

\subsection{Dusty Star Formation}

If the entire FIR emission in XID403 is interpreted as thermal emission from dust heated by young stars using the calibration derived by \citet{kennicutt98}, 
then a SFR of $\sim1000\;M_{\odot}$yr$^{-1}$ is derived. Such a large rate is not observed in the local Universe, but is frequently observed 
in high redshift galaxies (e.g. \citealt{magnelli12}). In the most luminous IR systems, the SFR derived from the FIR emission is usually larger than what is 
derived by the UV stellar light (e.g. \citealt{bongiorno12,casey13}). This is the case also for XID403, where the UV-estimated SFR, even after correcting for 
optical reddening, is a factor of $\sim3$
lower than that estimated from the IR. A plausible interpretation for this difference is that, in luminous IR systems, most stars form in very dusty regions 
and hence do not produce any light that can contribute to the optical SED. Recently, \citet{lofaro13} suggested that  the majority of the IR emission 
in LIRGs and ULIRGs (at $z\sim1-2$) is produced by diffuse dust (``cirrus'')  heated by intermediate-age ($10^8-10^9$yr) rather than by young 
($<10^8$yr) stellar populations. This would lead to an overestimate of the IR-derived SFRs by a factor of 2-3, while those derived from the UV-light would 
be more reliable indicators. However, for extreme star formation rates, such as that measured in XID403, the emission of dust heated by intermediate-age 
stars is expected to be small, and the total FIR emission appears to be a good proxy of the global SFR (see \citealt{lofaro13} for details). 

By fitting the FIR SED with a modified blackbody we find that the dust is warm,  $T_d=58.5\pm5.3$ K. 
Warm dust appears to be typical of ULIRGs both in the local (e.g. Arp220, \citealt{rangwala11}) and in the distant Universe 
\citep{fu12,riechers13}. This temperature has been derived by adopting a general grey body emission, i.e.
we did not assume that dust is optically thin. Should we adopt the optically-thin greybody approximation, as often done in the literature, 
we would measure a colder dust temperature $T_d\sim33$~K. This is because of the stronger emission at shortest wavelengths in the optically thin 
approximation (see Eqs. 1 and 2), which therefore needs a cooler temperature
to fit an SED bump at a fixed wavelength. This has been already discussed in the literature \citep{conley11,fu12}. We note here that the temperature
derived for XID403 in the optically-thin approximation is of the same order of that observed in ULIRGs and SMGs when using the same approximation 
\citep{casey12}. That is, the FIR peak emission of XID403 again seems in line with what has been observed in intense starburst systems. 

A different parameterization to the FIR SED on SMGs in the ALMA-LABOCA survey of the ECDFS (ALESS) has been adopted by \citet{swinbank13}.
They used three dust components (cold, $T=20-30$ K; warm, $T=50-60$ K; and hot, $T=80-120$ K) and allowed the dust emissivity index to 
vary within $\beta=1.5-2.0$ to derive mass and temperature of the various components: the cold and warm
components are those contributing to the peak SED emission, with the cold component dominating the total mass budget of the dust. Although the different
parameterizations prevent a straigthforward comparison between our results and those by \citet{swinbank13}, we note that the dust mass we derived 
is in line with their average value of $\sim4\times10^{8}\;M_{\odot}$. The  $L_{IR}$ and SFR we measured for XID403 are also comparable to those of high-z 
sources in the ALESS sample.

\citet{magdis12} studied the decrease of the gas-to-dust mass ratio with metallicity for a few samples of ``main-sequence'' and starburst galaxies
both in the local and distant Universe: starburst systems tend to have higher metallicities and lower gas-to-dust mass ratios than main-sequence star forming
galaxies. They then used the observed $M_{gas}/M_{dust}\propto 1/Z$ relation as a proxy to determine the molecular gas mass $M_{H_2}$ in their systems
($M_{gas}=M_{HI}+M_{H_2}$, so if both $M_{HI}, M_{dust}$, and $Z$ are known, $M_{H_2}$ can be derived) and finally estimate
the conversion factor $\alpha_{CO}$ between CO luminosity $L'_{CO}$ and molecular gas mass ($M_{H_2}=\alpha_{CO}L'_{CO}$). For local ULIRGs and distant
SMGs they found $\alpha_{CO}\sim1$, which is consistent with the standard value of 0.8 assumed for starburst systems \citep{solomon05}.
For XID403 we measured a dust mass of $M_d=4.9\pm0.8\times10^8\;M_{\odot}$. This, when compared with the total (atomic + molecular) gas
mass estimated from previous works (see Table 2), translates into a gas-to-dust mass ratio of $\sim50$ for a medium with solar metallicities \citep{nagao12},
and places XID403 in the region populated by local and distant starburst galaxies. The molecular gas mass derived using the relation by \citet{magdis12} is
$M_{H_2}\sim3.2\times10^{10}\;M_{\odot}$, suggesting that a conversion factor of the order of 0.8-1 also applies to XID403 
(\citealt{coppin10} indeed used $\alpha_{co}=0.8$ to derive $M_{H_2}=2.6\times10^{10}\;M_{\odot}$).

Our ALMA observations have placed a lower limit of $26\;M_{\odot}$yr$^{-1}$kpc$^{-2}$ to the star formation rate density in XID403. Assuming that both
atomic and molecular gas are co-spatial with the dust probed by ALMA observations, we can similarly estimate a gas surface density of 
$\Sigma_{gas}=\Sigma_{HI+H_2}=(M_{HI+H_2}/2)/(\pi r^2_{half})>6.6\times10^{8}\;M_{\odot}$kpc$^{-2}$ (see Table~2). These values place XID403 above the sequence 
of star forming galaxies following the Schmidt-Kennicutt law \citep{schmidt59,kennicutt98} that relates the surface density of star formation with that of 
gas observed in both local and distant 
disk galaxies \citep{daddi10}. Instead, XID403 falls within the sequence of starburst galaxies at $z=0-2$ of \citet{daddi10} , which are forming stars at a rate 
ten times higher than that of disk galaxies for a given gas density. 

\subsection{Physics of the system and blow-out condition} 

The high star formation density measured in XID403 is in line with the values measured in many distant star forming galaxies, which can even reach
values as high as $\sim3000\;M_{\odot}$yr$^{-1}$kpc$^{-2}$ \citep{diamond12,geach13,riechers13}. A ubiquitous feature
of star forming galaxies both in the local and distant Universe is that they undergo significant mass outflows \citep{bordoloi13}, 
with the outflow rate being proportional to the star formation rate density. In the extreme systems described by \citet{diamond12}, outflow velocities
of $\gtrsim1000$ km/s have been measured. 

It is then instructive to estimate the physical conditions in XID403 and provide an order of magnitude estimate about the likelihood that a large scale
outflow is launched. Assuming that a fraction $\epsilon$ of the available gas in the system is converted into stars (i.e. $\epsilon$ is the star formation efficiency), 
one has that $SFR=\epsilon f_{gas} M_{tot}/t_d$, where $M_{tot}$ is the total mass of the system, $f_{gas}$ is the gas fraction and $t_d$ is the dynamical time. 
By considering that the system velocity dispersion is $\sigma_{tot}\sim(GM_{tot}/R)^{1/2}$, where $R$ is the size of the system, and
that $t_d\sim(G\rho_{tot})^{-1/2}$, where $\rho_{tot}$ is the total mass density, one can write: $SFR\sim\epsilon f_{gas}\sigma_{tot}^3/G$.
The turbulent pressure that is produced by star formation onto the ISM (either through SN feedback,  stellar winds, H II region expansion, cosmic rays; 
\citealt{ostriker11}) is $P_t=\rho_{gas} \sigma_{gas}^2$, where $\sigma_{gas}$ is the turbulent velocity dispersion of the gas. It has been shown \citep{forster09,newman13} that moving from the
local Universe to $z\sim1-2$ the average gas velocity dispersion of star forming galaxies increases and is a strong function of the galaxy size. In particular, 
compact systems are more likely to be dispersion-dominated, with $\sigma_{gas}\sim70-80$ km/s for $r_{half}<1$ kpc at $z\sim1-2$. By extrapolating this trend
to $z>4$ (as also expected by models of galaxy growth, in which high-redshift systems are predominantly unstable, 
dispersion-dominated systems that later form extended rotationally supported disks; see \citealt{law12} and references therein), 
we will then assume that in XID403 the gas velocity 
dispersion is even higher than that of similar-size systems at $z\sim1-2$, and is of the same order of the total (stellar) velocity dispersion 
($\sigma_{gas}\approx\sigma_{tot}$).
We can therefore write the turbulent pressure injected by star formation into the ISM as: $P_t=\rho_{gas}[G\;SFR /(\epsilon f_{gas})]^{2/3}$. 
The total vertical weight in a self-gravitating gas disk is $P_{disk}=\pi G \Sigma_{gas}^2/2$, where $\Sigma_{gas}$ is the gas surface density \citep{ostriker11}. 
The conditions for a blow-out of the disk are met and a large-scale outflow is expected to occur when:

\begin{equation}
C \equiv P_t/P_{disk} = \frac{2}{\pi G^{1/3}}\frac{\rho_{gas}}{\Sigma^2_{gas}}\left(\frac{SFR}{\epsilon f_{gas}}\right)^{2/3} >1.
\label{cblow}
\end{equation}

To estimate $\rho_{gas}$, assuming the gas is distributed into a disk, its scale height needs to be known. The detailed morphology of XID403 is unknown.
Starbursting systems at high redshift may have complex morphologies, possibly as a result of recent mergers \citep{schinnerer08,tacconi08}. 
However, disk-like morphologies of cold gas have been observed in a significant fraction of starforming galaxies hosting QSOs at epochs as early as 
$z\sim6$ \citep{wang13}. By assuming a scale-height of the order of the disk size (i.e. a spherical volume), we can at least provide a lower limit 
to $\rho_{gas}$, an upper limit to $\epsilon \propto t_d\propto \rho_{gas}^{-1/2}$, and hence a conservative lower limit to $C$.
Then, for $r_{half}<2.5$ kpc, $\rho_{gas}>1.3\times10^{-23}$ g cm$^{-3}$, $\tau_{d}<3.4\times 10^7$yr, $\epsilon<1$. Finally, by considering 
$f_{gas}=M_{gas}/(M_{gas}+M_*)\sim0.2$ and substituting into Eq.~\ref{cblow}, we obtain $C>5$, suggesting that the blow-out condition is met
and a large scale gas outflow should occur. Interestingly, this computed blow-out condition is scale-free. Indeed, 
$C\propto \rho_{gas} \Sigma_{gas}^{-2}\tau_d^{-2/3}\propto r^{-3} r^4 r^{-1}\propto r^0$. Therefore, the blow-out condition will be met also for
a system size of $r_{half}\sim0.9$ kpc or smaller (note that this blow-out condition cannot be applied to large, $>>2$kpc, rotation-dominated systems).

\subsection{Possible evidence for outflowing gas}

\begin{figure}
\resizebox{\hsize}{!}{\includegraphics[angle=0]{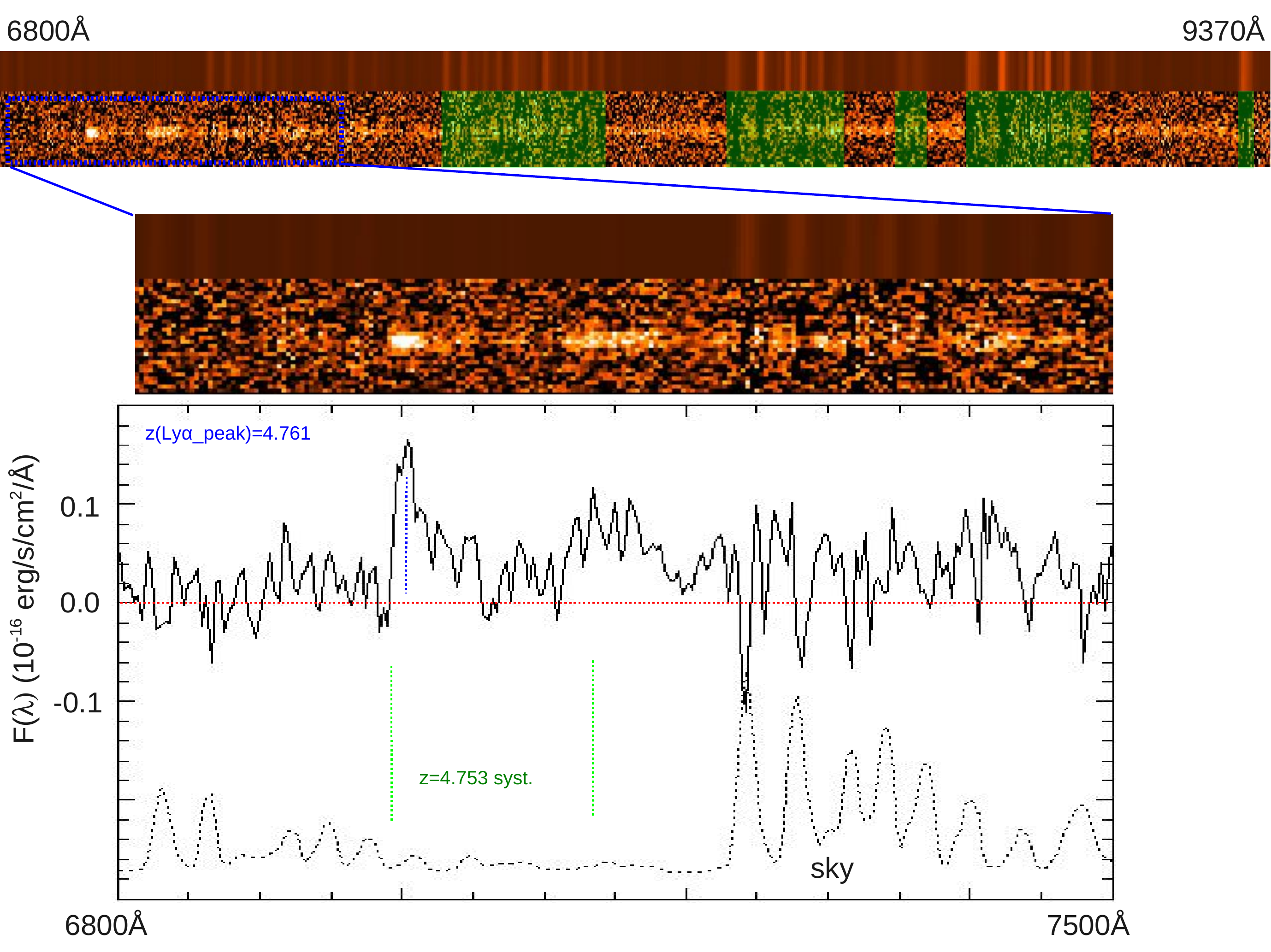}}
\caption{The FORS2 two-dimensional spectrum is shown from 6800 to 9400\AA~(top).
The shaded green rectangles mark the noisy regions affected by intense sky emission lines.
A zoom on the Ly$\alpha$ and N~{\sc v} positions in the wavelength interval 6800-7500\AA~(blue dotted rectangle)
is shown in the middle and bottom of the figure. The wavelength of the Ly$\alpha$ and N~{\sc v} at the
systemic redshift derived from the [NII]205$\mu$m line \citep{nagao12} are also reported (green vertical dotted lines).}
\label{eros}
\end{figure}

Based on the above considerations we searched for evidence of an outflow in XID403, and first considered its optical spectrum.
Besides the observations performed at Keck \citep{coppin09}, the source was observed twice during the GOODS-FORS2 campaign \citep{vanzella08}.
We improved the data reduction originally published in \citet{vanzella08} by including the more recent procedures adopted in \citet{vanzella11}.
The two FORS2 spectra were re-analyzed and stacked, providing a final spectrum of 6.3 hours integration time (see Fig.~\ref{eros}).

A first piece of evidence for outflowing gas could be the velocity shift observed between the Ly$\alpha$ emission and the sub-mm lines of [NII], [CII], CO(2-1)
\footnote{\citet{coppin10} also report a source redshift measurement of $z=4.751\pm0.005$ derived from a [OII]3727 emission line in a near-IR 
VLT spectrum, that is consistent with what is measured from the submm lines.}
(see Fig. 2 in \citealt{nagao12} and Fig.~\ref{eros}). If one assumes that e.g. the [NII] emission is tracing the systemic redshift, 
then Ly$\alpha$ is red-shifted by $\sim$ 350 km~s$^{-1}$.
Redshifted Ly$\alpha$ emission by 100-1000 km~s$^{-1}$ is commonly observed in Lyman Break Galaxies \citep{steidel10}, and is interpreted as evidence for
outflowing gas: in this scenario the observed  Ly$\alpha$ emission would  arise from the receding side of the outflow, once the Ly$\alpha$ photons 
have been redshifted by several hundreds km~s$^{-1}$ relative to the bulk of the material they have to cross to reach us (blue-shifted Ly$\alpha$ photons are instead efficiently absorbed within the system). This interpretation is however complicated by the fact that radiative transfer processes 
strongly shape the emerging  Ly$\alpha$ line profile and peak wavelength, which in fact depend on the gas column density, dust mass and their unknown 
geometrical distribution. Moreover, at high redshift, the blue side of the Ly$\alpha$ emission can be also depressed 
by absorption in the neutral intergalactic medium (IGM). 
Besides Ly$\alpha$ emission, the optical spectrum of XID403 also shows a rather broad ($\sim 2200$ km~s$^{-1}$ FWHM) N~{\sc v} 1240 feature of 
comparable flux, which was 
regarded as the signature of an AGN \citep{vanzella06,coppin09}. One possible problem with the AGN interpretation 
is the width of the N~{\sc v} line, which should be produced by Broad Line Clouds 
at sub-pc scales from the black hole, and would suggest an unobscured view of the nucleus (unless one is seeing scattered radiation): such a direct-view  
scenario is probably unfavored by the large amount of obscuration measured in the X-rays and also by the large amount of gas and dust
confined within a radius that could be as small as a few hundreds pc (see Section 7.2). Here we suggest that, instead, the broad N~{\sc v} 1240 could arise
from stellar winds produced by hot (O) stars present in this massive starburst system. These winds would produce P-cygni profiles. Although, to our knowledge, no
such features have been observed at high redshift, they have indeed been detected in starforming galaxies in the local Universe \citep{heckman11,leitherer13}. 
In particular, \citet{heckman11} present  
HST/COS UV spectra of a sample of ten Lyman Break Analogs, i.e. a rare population of local galaxies with UV luminosities and surface densities 
similar to high-z Lyman Break Galaxies, where high velocity outflows have been detected ($\sim 600$ km~s$^{-1}$ median) in all of them and very compact 
starbursts ($r\sim0.1$ kpc) in about 40\% of them. The UV spectra presented by Heckman et al. (2010), are strikingly similar to that of XID403, i.e. they also have 
Ly$\alpha$ peaks redshifted with respect to the systemic velocity and similar Ly$\alpha$ profiles, Ly$\alpha$/N~{\sc v} ratios, N~{\sc v} widths. 
We also note that the profile of the N~{\sc v} line in XID403 may hint to a P-Cygni profile with a depression in the continuum blueward of N~{\sc v} 
(see Fig.~\ref{eros}). The detection of the stellar UV continuum and line emission is  in agreement with the SED analysis 
(see Fig.~\ref{sed_ds}). 
In extremely dusty systems, the detection of stellar UV light can be ascribed to an inhomogeneous ISM, as suggested for instance by the analysis of 
\citet{leitherer13}  on four local UV-selected LIRGs. We therefore conclude that in XID403 there is reasonable evidence for strong stellar winds from hot and
massive stars. 
Together with supernovae explosions \citep{murray05sn} and AGN feedback \citep{maiolino12} these winds are one of those mechanisms which would be able 
to launch galactic-scale outflows \citep{leitherer92}.

Another emission feature of XID403 which deserves some discussion is the line at $\sim$6.9 keV likely arising from  Fe~{\sc xxvi}, 
for which a rest frame equivalent width of $EW_{rest}=2.8_{-1.4}^{+1.7}$keV was measured. The line detection is admittedly uncertain (2$\sigma$), 
but we note that prominent emission lines from highly-ionized iron have been found in many
LIRGs and ULIRGs (with or without AGN) both in the local and distant Universe \citep{iwasawa09,lindner12}.
Notable examples of such lines, with EW up to $\sim2$ keV, have been detected in powerful starforming objects which are known to contain heavily obscured AGN
like Arp220 \citep{iwasawa05}, the Superantennae \citep{jia12}, IRAS 12127-1412  \citep{nardini11}
or the HyLIRG IRAS 00182-7112 \citep{nandra07}. The origin of the line in these systems is far from being understood. 
\citet{matt96} suggested that in obscured AGN, features from highy ionized iron (ionization parameters $\xi=10^3-10^4$) could be produced by the 
so-colled ``warm mirror'', i.e. an optically thin plasma located on pc scales around the black hole. In local heavily obscured AGN, 
however, emission lines at 6.7-6.9 keV with EW$\sim$1-2 keV are far less common than the line at 6.4 keV, which is explained as fluorescent 
iron emission by an almost neutral reprocessor on the same pc scales, and are always accompanied by, and weaker than, the cold 6.4 keV line
(in XID403 only loose constraints can be placed to the 6.4 keV line: EW$\lesssim$4 keV; \citealt{gilli11}).\footnote{We note that the line at $\sim6.9$ keV 
rest-frame is unlikely to be a blue-shifted line at 6.4 or 6.7 keV rest-frame from Fe~{$<$\sc xxvi}, which would imply outflow 
velocities of $\sim23,000$ or $\sim9,000$km $s^{-1}$, respectively, i.e. a factor of $\sim20-50$ larger than what is estimated
from Ly$\alpha$.  While such ultra fast outflows (UFOs; e.g. \citealt{tombesi13}) of ionized gas have been seen in powerful AGN as absorption lines, 
they have never been observed in emission.}

\citet{nandra07} suggested that the $EW\sim1$ keV line at 6.7 keV seen in the $z=0.3$ HyLIRG IRAS 00182-7112 might originate from a thin 
plasma which would be difficult to confine and should be then outflowing. Both the hidden QSOs in IRAS 00182-7112 and XID403 are indeed powerful 
enough to launch a plasma outflow driven by radiation pressure \citep{tombesi13}. Also, the luminosity of the Fe line in XID403 is of the same order of the 
X-ray luminosity expected for a 1000 $M_{\odot}$yr$^{-1}$ starburst (assuming the SFR vs $L_x$ relations by \citealt{ranalli03}), and it is therefore unlikely that the 
starburst can provide enough photons at $E>8$ keV to ionize iron atoms up to Fe~{\sc xxvi}. It is nonetheless possible that both supernovae explosions 
and stellar winds excavate channels through the ISM making this plasma visible to the observer. This might explain why such strong lines from ionized iron 
are preferentially found in starburst galaxies.

If the detection of the $\sim6.9$keV line in XID403 is real and it is associated to outflowing plasma, we can roughly estimate its mass outflow rate

The line energy centred at $\sim6.9$ keV would imply a high ionization parameter $\xi\sim10^3-10^4$, where

\begin{equation}
\xi=L_{ion}/(n_eR^2)\; {\rm erg\;cm\;s}^{-1} ,
\label{xi}
\end{equation}

and $L_{ion}$ is the source ionizing luminosity integrated between 13.6 eV and 13.6 keV \citep{tarter69},
$R$ is the distance to the ionizing source, and $n_e$ is the average electron density of the outflow. 

In the case of a spherical or biconical geometry the outflow rate can be approximated with \citep{mckernan07}:
\begin{equation}
 \dot{M}_{out} \approx \mu_H n_e m_p \Omega R^2 v_{out} , 
\label{mdot}
\end{equation}

where $\mu_H$ is the mean atomic mass per hydrogen atom, $n_e$ is the average electron density of the outflowing plasma, $m_p$ is the proton mass, 
$\Omega$ is the solid angle subtended by the outflow ($\Omega=4\pi$ for a spherical outflow), 
$R$ is the radius of the outflow (distance to the nucleus), and $v_{out}$ the outflow velocity (here estimated through the Ly$\alpha$ shift). 
No clumpiness of the plasma is assumed.
Solving Eq.~\ref{xi} for $n_eR^2$ and substituting into Eq.~\ref{mdot}, returns:
\begin{equation}
 \dot{M}_{out} \approx \mu_H m_p \Omega v_{out} L_{ion}/\xi.
\label{mdot2}
\end{equation}

By assuming $\mu_H=1.3$ (valid for solar abundances) and $L_{ion}=L_{bol}/2$ (as measured from the average intrinsic QSO SED by \citealt{elvis94}) we obtained:
\begin{equation}
\dot{M}_{out} \approx 72 \left(\frac{\Omega}{4\pi}\right)\left(\frac{10^3}{\xi}\right)\left(\frac{v_{out}}{350}\right)\; M_{\odot}{\rm yr}^{-1}.
\end{equation}

The uncertanties on the estimated plasma outflow rate are admittedly large. In particular the rate would decrease for biconical outflows 
and ionization parameters $\xi>10^3$. However, one may conclude that from a few to a few tens of $M_{\odot}$ of  highly ionized gas are expelled
from the system every year. Obsevations of nearby AGN show that most of the outflowing mass is in a cool atomic or molecular phase rather 
than in a highly-ionized phase. For instance, \citet{tombesi12} measured plasma outflow rates of $<1M_{\odot}$~yr$^{-1}$ in a sample of local Seyfert 1 galaxies,
while outflow rates of $10-100\;M_{\odot}$~yr$^{-1}$ and $100-1000\;M_{\odot}$~yr$^{-1}$ have been observed for neutral \citep{rupke05} and molecular 
\citep{sturm11} gas, respectively, in local ULIRGs hosting AGN of similar bolometric luminosity. Then, if we assume that also in XID403 most of the outflowing 
gas is molecular, the total mass of gas that is expelled  may be of the same order of or even higher than that going into stars,
as it is often observed in star forming galaxies with \citep{feruglio10,maiolino12} or without \citep{newman12, bolatto13} the presence of an AGN .

\subsection{The descendants of XID403-like systems}

The high SFR (1020 $M_{\odot}$~yr$^{-1}$) and the possibility that the outflow rate is similar to the SFR, suggest that XID403 will be
depleted of all the available gas $M_{gas}\sim2.6\times10^{10}\;M_{\odot}$ in $t_{dep}=M_{gas}/(SFR+\dot{M}_{out})\approx 1.3\times10^7$ yr. 
Assuming that no further star formation would occur since then, XID403 would appear as an old, quiescent galaxy by $z\sim 3$, once its stellar populations
have evolved for  $\sim$1 Gyr. \citet{coppin09} showed that SMGs at $z\sim4-5$ possess the baryonic mass and gas depletion time-scales necessary to
be the progenitors of luminous quiescent galaxies at $z\sim3$, and that their space density is indeed consistent with that predicted by galaxy formation
models. Based on our high-resolution submm observations with ALMA, we here take a step further and suggest that objects like XID403 have also the
right size to be the progenitors of quiescent galaxies at $z\sim3$. Indeed, it has been shown \citep{daddi05,trujillo07,cimatti08,vandokkum08} that a 
substantial fraction 
of passive galaxies at redshifts $z>1$ have sizes significantly smaller (a factor of 3 to 5) than local passive galaxies of equal mass, featuring 
stellar densities as high as $\Sigma_*=(M_*/2)/(\pi r^2_{half})>10^{10}\;M_{\odot}$~kpc$^{-2}$ for ultra-compact systems \citep{cassata11}. 
One open question is how such old and compact remnants have formed.  Very recently \citet{barro13} identified a population of compact star forming 
galaxies (cSFGs) at $z=2-3$, whose number density, mass, size and star formation rates qualify them as likely progentitors of compact quiescent galaxies (cQGs) 
at $z\sim1-1.5$ (see also \citealt{williams13}). The incidence of nuclear activity in cSFGs is very high: about 30\% of them appear to host an X-ray luminous 
AGN ($L_x>10^{43}$ erg s$^{-1}$), while the AGN fraction in non-compact star forming galaxies with similar mass and redshift is less than 1\%. 
This suggests that AGN feedback may be responsible 
for a rapid evolution of star formation in cSFGs, leaving compact remnants on dynamical timescales ($\sim10^{8}$ yr). XID403 appears to be similar to the cSFGs 
discussed by \citet{barro13} and hence to satisfy all the requirements to be the progenitor of a cQG at $z\sim3$. 
The stellar density measured in XID403 (assuming $r_{half}$ = 1 kpc, as suggested by CANDELS data at 2800\AA\ rest-frame) 
is $\Sigma_*\sim 1.8\times 10^{10}\;M_{\odot}$~kpc$^{-2}$, comparable to  that of ultra-compact galaxies at $z>1$. 
At a rate of 1020 $M_{\odot}$~yr$^{-1}$, its measured stellar mass of $1.1\times10^{11}\;M_{\odot}$ must have been built in $\sim10^8$ yr, 
and it will grow at most by $\sim20\%$ ($M_{*, end} = M_*+M_{gas} = 1.23 M_*$) in the next $\sim10^7$ yr. 
This shows that most of the compact stellar core in XID403 is already in place. In addition, our ALMA data indicate that the vigorous star formation 
that is building that core $is$ indeed happening on those sub-kpc scales, perhaps as a result of a major merger event \citep{hopkins08} and/or 
direct feedback (first positive and then negative) from the hidden QSO \citep{zubovas13}.

\section{Conclusions}

We have reported ALMA Cycle 0 observations of the ULIRG XID403 at $z=4.75$ in the Chandra Deep Field South (CDFS). This system hosts the most 
distant Compton-thick QSO known to date and is an excellent laboratory to study the coevolution of a supermassive black hole with its host galaxy in the early 
Universe.  We have complemented our sub-mm photometry with other data from ALMA and Herschel and built the far-IR SED of XID403. In addition, we took 
advantage of the dense UV to mid-IR coverage of the CDFS, and in particular of the CANDELS database, to build a broad band SED and obtain a spectral 
decomposition between the QSO SED and that of its host. We finally searched evidence of outflowing gas and re-analyzed the Chandra X-ray spectrum 
looking for Fe emission features associated to it. 
Our main results are the following:

$\bullet$
The source emission at 1.3 mm does not appear to be resolved  in our ALMA data at $\sim0.75$ arcsec resolution. This places an upper limit of 2.5 kpc to the 
half-light radius of the continuum emission from dust heated by star formation. After deconvolving for the beam size, however, we found a $\sim3\sigma$ 
indication of an intrinsic source size of $0.27\pm0.08$ arcsec (Gaussian FWHM), which would correspond to $r_{half}\sim0.9\pm0.3$ kpc. Further observations
with ALMA at even higher resolution would fully resolve the source emission and hence accurately determine its physical size and properties.

$\bullet$
By fitting the far-IR SED with a modified blackbody spectrum we measured a warm dust temperature, $T_d=58.5\pm5.3$ K, that is comparable 
to what has been observed in other high-z submillimeter galaxies. The star formation rate derived from the far-IR SED is 
SFR=$1020\pm150\;M_{\odot}$ yr$^{-1}$,  in agreement with previous estimates at lower S/N. Based on the measured SFR and source size, we constrain the 
SFR surface density to be $\Sigma_{SFR}>26\;M_{\odot}$yr$^{-1}$kpc$^{-2}$ ($\Sigma_{SFR}=200\;M_{\odot}$yr$^{-1}$kpc$^{-2}$ for $r_{half}\sim0.9$ kpc), 
similar to what is observed in other local and high-z starburst galaxies.

$\bullet$
In the plausible assumption that both the molecular and atomic gas masses - derived from  previous [CII] and CO(2-1) observations at low angular resolution - 
are co-spatial with dust and assuming $r_{half}\sim0.9\pm0.3$ kpc, we derived a column density of $N_H\sim0.3-1.1\times10^{24}$cm$^{-2}$ 
towards the central SMBH. This is consistent with the column density of $1.4^{+0.9}_{-0.5}\times10^{24}$cm$^{-2}$ measured from the X-rays. Therefore, in principle, 
if  both gas and dust were confined within sub-kpc scales, this would be sufficient to produce the observed X-ray column density without any need of a 
pc-scale absorber (e.g. the torus postulated by Unified Models).

$\bullet$
We fitted the broad band  SED of XID403 with a composite spectrum made of dusty simple stellar populations (SSPs) plus AGN emission reprocessed by hot dust.
The optical and far-IR parts of the SED are well modeled by the direct and dust-reprocessed emission of stars, respectively. Hot-dust heated by the AGN is
needed to fit the data at mid-IR wavelengths (4$\mu$m rest-frame). We measured a total stellar mass of $1.1\times10^{11}\;M_{*}$ and an AGN bolometric 
luminosity of $\sim10^{46}$ erg/s, which is about half of that produced by stars.

$\bullet$
We argued that most if not all of the UV/optical light in XID403 is produced by stars. On the one hand, an excellent fit to the smooth UV/optical SED is obtained
just with stellar emission, without any need of an additional component. On the other hand, both the Ly$\alpha$ and N~{\sc v} emission features in the UV/optical
spectrum can be ascribed to the stong starburst. In particular, the N~{\sc v} line, which is often regarded as an AGN signature, is in this case likely 
arising from stellar winds produced by hot and young (O) stars, as seen in local compact starbursts. If the AGN contribution to the UV/optical light is then 
negligibe, the unresolved morphology in the HST/ACS z-band data would suggest that half of the stars could be located in a region even smaller than 
0.3 kpc radius, providing an even stronger constraint to the system size.

$\bullet$
We speculated that the high compactness of star formation, together with the presence of a powerful AGN, likely produce an outflowing wind. 
This would be consistent with the $\sim350$ km~s$^{-1}$ velocity shift observed between the Ly$\alpha$ emission and the submm lines ([CII], CO(2-1), [NII]) 
and with the highly-ionized Fe emission line at $\sim6.9$ keV rest-frame tentatively observed in the X-ray spectrum. The forthcoming Chandra 3Ms exposure 
of the CDFS will almost double the X-ray photon statistics in XID403. This will provide an excellent opportunity to test the reality of the detected iron feature 
and constraint at best all physical parameters extracted from the current X-ray spectrum.

$\bullet$
Our analysis showed that a compact, potentially sub-kpc, stellar core,  is already in place in XID403, with a stellar density similar to that of ultra-compact 
early type galaxies observed at $z>1$. In particular, our high-resolution data from ALMA showed that the vigorous star formation that is building that core 
is indeed 
happening on the same sub-kpc scales. Besides the mass, star formation rate and gas depletion timescales, XID403 has then also the right size to be one 
of the progenitors of the compact quiescent massive galaxies seen at $z\sim3$.

\begin{acknowledgements}
We thank Carlos De Breuck for sharing with us some of his unpublished ALMA results on XID403. The anonymous referee is acknowledged for useful comments.
We acknowledge support from the Italian Space Agency under the ASI-INAF contract I/009/10/0 and from INAF under the contract PRIN-INAF-2012.
FC acknowledges financial support from PRIN MIUR 2010-2011,
project "The Chemical and Dynamical Evolution of the Milky Way and Local Group Galaxies", prot. 2010LY5N2T.       
This paper makes use of the following ALMA data: ADS/JAO.ALMA\#2011.0.00716.S. ALMA is a partnership of ESO (representing its member states), NSF (USA) and NINS (Japan), together with NRC (Canada) and NSC and ASIAA (Taiwan), in cooperation with the Republic of Chile. The Joint ALMA Observatory is operated by ESO, AUI/NRAO and NAOJ.
\end{acknowledgements}


\bibliographystyle{aa} 
\bibliography{../../../neon/biblio} 

\end{document}